\documentstyle[11pt,newpasp,twoside,epsf]{article}
\def\edcomment#1{\iffalse\marginpar{\raggedright\sl#1\/}\else\relax\fi}
\reversemarginpar

\begin{document}
\title{The Role of Clouds in Brown Dwarf and Extrasolar Giant Planet
Atmospheres}
\author{Mark S. Marley}
\affil{NASA Ames Research Center; Mail Stop 245-3; Moffett Field, CA
and New Mexico State University; Department of Astronomy; Las Cruces, NM 88003}
\author{Andrew S. Ackerman}
\affil{NASA Ames Research Center; Mail Stop 245-4; Moffett Field, CA
94035}

\begin{abstract} Clouds and hazes are important throughout our
solar system and in the atmospheres of brown dwarfs and extrasolar
giant planets. Among the brown dwarfs, clouds control the colors
and spectra of the L-dwarfs; the disappearance of clouds
helps herald the arrival of the T-dwarfs.  The structure and
composition of clouds will be among the first remote-sensing
results from the direct detection of extrasolar giant planets.
\end{abstract}

\section{Introduction}

Even before the first discovery of brown dwarfs and extrasolar 
giant planets (EGPs) it 
had been apparent that a detailed appreciation of cloud
physics would be required to understand the atmospheres of these objects
(e.g. Lunine et al. 1989).
Depending on the atmospheric effective temperature, Fe, $\rm Mg_2SiO_4$,
$\rm MgSiO_3$, $\rm H_2O$, and $\rm NH_3$ among others may condense 
in substellar atmospheres.  Since
every atmosphere in the solar system is 
influenced by clouds, dust, or hazes, the need to follow the fate of 
condensates in brown dwarf and EGP atmospheres is self-evident.
What has become clearer over the past five years is that details such as
the vertical structure and particle sizes in clouds play a decisive role
in controlling the thermal structure and emergent spectra from these
atmospheres.  Indeed the available data are already sufficient to 
help us choose among competing models.

In this contribution we will briefly summarize some of the roles clouds play in
a few solar system atmospheres to illustrate what might
be expected of brown dwarf and extrasolar giant planet atmospheres.  Then
we will summarize a new cloud model developed to study these
effects, present some model results, and compare them to data.  Since
brown dwarfs have similar compositions and effective temperatures
to EGPs  and a rich dataset already exists, we focus on the lessons 
learned from the L- and T-dwarfs.  We then
briefly review the importance of clouds to EGP atmospheres 
and future observations.

\section{Clouds in the Solar System }

Clouds dramatically alter the appearance, thermal structure, and
even evolution of planets.
Venus glistens white in the morning and evening skies because sunlight
reflects off of its bright cloud tops.  If there were no condensates in Venus'
atmosphere  the planet would take on a bluish hue from Rayleigh
scattered sunlight.  Mars' atmosphere is warmer 
 than it would otherwise be thanks to absorption of incident solar radiation
by atmospheric dust (Pollack et al. 1979).  
The effectiveness of the $\rm CO_2$ greenhouse during Mars's
putative warm and wet early history is tied to poorly understood
details of its cloud physics and radiative transfer (Mischna, et al. 2000).
Indeed the future climate of Earth in a fossil-fuel-fired greenhouse may
hinge on the role water clouds will play in altering Earth's albedo
and scattering or absorbing thermal radiation.

The appearance of the Jovian planets is controlled by
the extensive cloud decks covering their disks.  On Jupiter and Saturn 
thick $\rm NH_3$ clouds, contaminated by an unknown additional
absorber, reflect about 35\% of incident radiation back to space.
$\rm  CH_4$ and $\rm H_2S$ clouds play a similar role at Uranus and Neptune.

\begin{figure}
\plotone{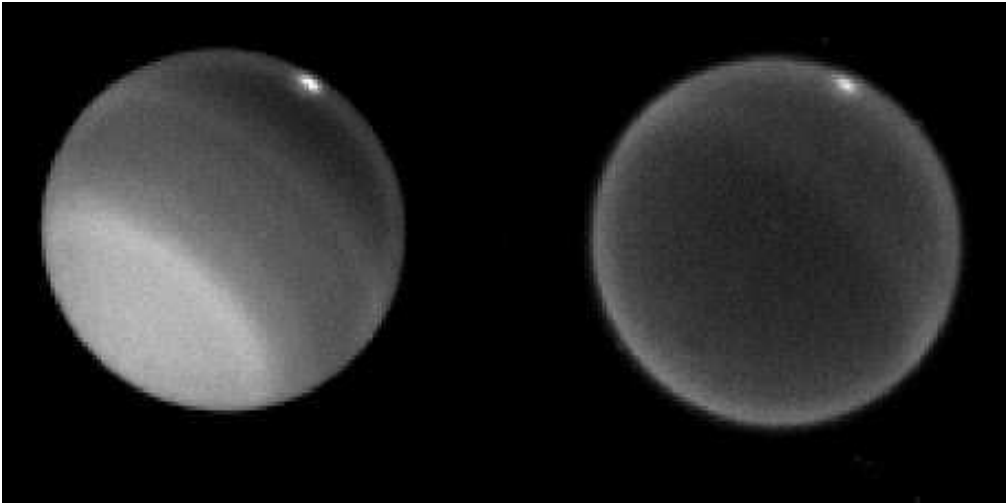}
\caption{Near consecutive HST images of Uranus taken 
through different filters. The filter employed for the left hand image
probes a broad spectral range from $0.85$ to $1\,\rm \mu m$
while the right hand image is taken through a narrow filter 
sensitive to the $0.89\,\rm \mu m$ $\rm CH_4$ absorption band.  
The relative visibility of various cloud features between the 
two images is a measure of the cloud height as the incident
photon penetration depth is modulated by methane absorption.  Images
courtesy H. Hammel and K. Rages.}
\end{figure}

The vertical structure of the jovian cloud layers was deduced 
by variation of their reflected spectra inside and outside of
molecular absorption bands.
Figure 1 illustrates this process.  In the left hand
image incident sunlight 
penetrates relatively deeply into the atmosphere and is scattered 
principally by a
cloud deck over the south pole and a bright cloud near the northern
mid-latitude limb.  The relative heights of these two features cannot
be discerned from this single image.  The right hand image, however,
was taken in the strong $0.89$-$\rm \mu m$ methane absorption band.  Here
the south polar cloud is invisible since incident sunlight
is absorbed by $\rm CH_4$ gas above the cloud before it can scatter.  We
conclude that the bright northern cloud lies higher in the atmosphere
since it is still visible in this image.  The application of this
technique to spectra and images of the giant planets has yielded 
virtually all the information we have about the vertical
structure of these atmospheres (e.g. West, Strobel, \& Tomasko  1986; 
Baines \& Hammel 1994; Baines et al. 1995).
A similar reasoning process can be applied to brown dwarf and 
EGP atmospheres.

The large body of work on jovian clouds cannot be easily generalized, but two
robust results are apparent.  First, sedimentation of cloud droplets
is important.  Cloud particles condense from the atmosphere, coagulate,
and fall. The fall velocity depends on the size of the drops and
the upward velocity induced by convection or other motions in the
atmosphere. They do not stay
put.  A diagnostic often retrieved from imaging or spectroscopic
observations of clouds is the ratio of the cloud particle scale height
to that of the gas.  If condensates were distributed uniformly
vertically in the atmosphere this ratio would be 1.  Instead numerous
investigations have found a ratio for Jupiter's ammonia clouds of 
about 0.3 (Carlson, Lacis, \& Rossow 1994).
The clouds are thus relatively thin in vertical extent.  
The importance of sedimentation is borne out even for unseen 
Fe clouds, for example, by Jupiter's
atmospheric chemistry (Fegley \& Lodders 1994).

A second important result is that cloud particles are large, a result
of coagulation processes within the atmosphere.  Sizes are difficult to
infer remotely and the sizes to which a given observation is sensitive
depend upon the wavelength observed.  Nevertheless it is clear that
Jupiter's ammonia clouds include particles with radii exceeding 1 to 
$10\,\rm \mu m$, much larger than might be expected simply by direct
condensation from vapor in the presence of abundant
condensation nuclei (Carlson et al. 1994; Brooke et al. 1996).  
Similar results are found for ammonia clouds on Saturn (Tomasko et al. 1984)
and methane clouds in Uranus and Neptune (Baines et al. 1995).

These two lessons from the solar jovian atmospheres -- clouds have
finite vertical extents governed by sedimentation and large
condensate sizes -- guide us as we consider clouds in brown dwarf
and extrasolar giant planet atmospheres.
 
\section{Evidence of Clouds in Brown Dwarf Atmospheres}

The first models of the prototypical T-dwarf Gl 229 B established that grains
play a minor role, if any, in controlling the spectrum of
the object.  The early Gl 229 B models of  Marley et al. (1996), Allard
et al. (1996) and Tsuji et al. (1996) all found  best fits
to the observed spectrum by neglecting grain opacity.  This provided
strong evidence that any cloud layer was confined below the
visible atmosphere.  All the models, however, shared the same shortcoming
of predicting infrared water bands deeper than observed.
Another difficulty with the early models is that they either predicted
too much flux shortwards of $1\,\rm \mu m$ (Marley at al. 1996) or used
unrealistic molecular opacities (Allard et al. 1996) to lower the optical flux.
Griffith, Yelle, \& Marley (1999) and Tsuji, Ohnaka, \& Aoki (1999) suggested
variations of particulate opacity to lower the flux, but ultimately
Burrows, Marley, \& Sharp (2000) argued that broadened alkali metal
bands were responsible for the diminution in flux, a prediction
verified by Liebert et al. (2000). 

\begin{figure}
\plotone{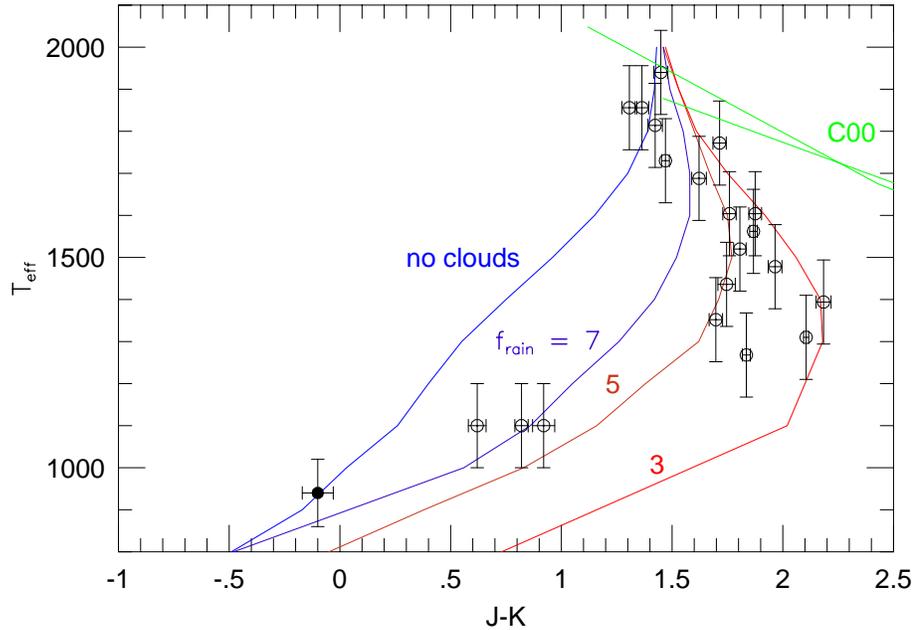}
\caption{$J-K$ color of brown dwarfs as a function of
$T_{\rm eff}$.  
Open datapoints represent L- and T-dwarf colors measured by
Stephens et al. (2001) with L-dwarf temperatures estimated from fits of
$K-L^\prime$ to models of Marley et al. (2001).  Since $K-L'$ is relatively
insensitive to the presence or absence of clouds for the L-dwarfs
it provides a good $T_{\rm eff}$ scale (Marley et al. 2001).
The early T-types ($0.5 < J-K < 1$) are arbitrarily all assigned 
to $T_{\rm eff} = 1100\,\rm K$.  Likewise model $T_{\rm eff}$s are given
estimated error bars of $\pm 100\,\rm K$.  The filled circle represents the position
of the prototypical T-dwarf Gl 229 B
(Saumon et al. 2000; Leggett et al. 1999).
Four model cases are shown from the work of Marley
et al. (2001): evolution with no clouds, and with clouds following
the prescription of Ackerman \& Marley (2001) with $f_{\rm rain} $
(rainfall efficiency, see text)
varying from 7 (heavy rainfall) to 3 (moderate rain).  Also shown
are colors (C00) from models by Chabrier et al. (2000) in which there is
no downward transport of condensate.  The Marley et al. model lines are for objects
with gravity $g=1000\,\rm m\,sec^{-2}$, roughly appropriate for a
$30\,\rm M_J$ object. There is little dependence of $J-K$ on
gravity in this regime. The Chabrier et al. lines are for 30 and
$60 \,\rm M_J$ objects.  }
\end{figure}

The first confirmation that dust was present in the atmospheres of at least
some brown dwarfs came with the discovery of the warmer L-dwarfs.  These
objects, unlike the methane-dominated T-dwarfs, have red colors
in $J-K$ and spectra that have been best fit with dusty atmosphere models
(Jones \& Tsuji 1997), although a complete analysis does not yet exist.
The difficulty arose in explaining how the dusty, red
L-dwarfs evolved into the clear, blue T-dwarfs (Figure 2).  
Models in which dust does not settle into discrete cloud layers (Chabrier
et al. 2000) 
predict that cooling brown dwarfs would become redder 
in $J-K$ with falling effective temperature as more and more
dust dominates the atmosphere.  Since the atmosphere models employed
in this work ignore the lessons learned from our jovian planets 
(they employ sub-micron particle sizes 
and do not allow the dust to settle) it is not surprising that they
do not fit the data.

\section{A New Cloud Model}

A number of models have been developed to describe the cloud formation
processes in giant planet and brown dwarf atmospheres.  Ackerman \& Marley
(2001) describe these in some detail.  In general these models
suffer from a number of drawbacks which limit their utility
for brown dwarf and EGP modeling.  Some rely upon free parameters 
which are almost impossible to predict while others do not
predict quantities relevant to radiative transfer of in the atmosphere.  
For example, the atmospheric supersaturation cannot be specified
without a detailed knowledge of the number of condensation nuclei available. 
Ackerman \& Marley  developed
a new eddy sedimentation model for cloud formation in substellar
atmospheres that attempts to predict cloud particle sizes and vertical
extents.

Ackerman \& Marley argue that in terrestrial clouds the downward
transport of large drops as rain removes substantial mass from clouds
and reduces their optical depth.  Yet properly modeling the condensation,
coagulation, and transport of such drops requires a complex microphysical 
model and a concomitant abundance of free parameters.  In an
attempt to account for the expected effects of such microphysical processes 
without modeling them in detail,
they introduce a new term into the equation 
governing the mass fraction $q_t$ of an atmospheric condensate at a given
altitude $z$ in an atmosphere:
$$K{{\partial q_t}\over{\partial z}} + f_{\rm rain} w_* q_c = 0. \eqno(1)$$
Here the upward transport of the vapor and condensate
is by eddy diffusion
as parameterized by an eddy diffusion coefficient $K$.  In equilibrium
this upward transport is balanced by the downward transport of 
condensate $q_c$.  
The free parameter $f_{\rm rain}$ has been introduced as the ratio of 
the mass-weighted droplet sedimentation velocity to $w_*$, the
convective velocity scale.  In essence $f_{\rm rain}$
allows downward mass transport to be dominated by massive drops
larger than the scale set by the local eddy updraft velocity: in other
words, rain.  Ackerman \& Marley (2001) treat $f_{\rm rain}$ as
an adjustable parameter and explore its consequences.  

\begin{figure}
\plotone{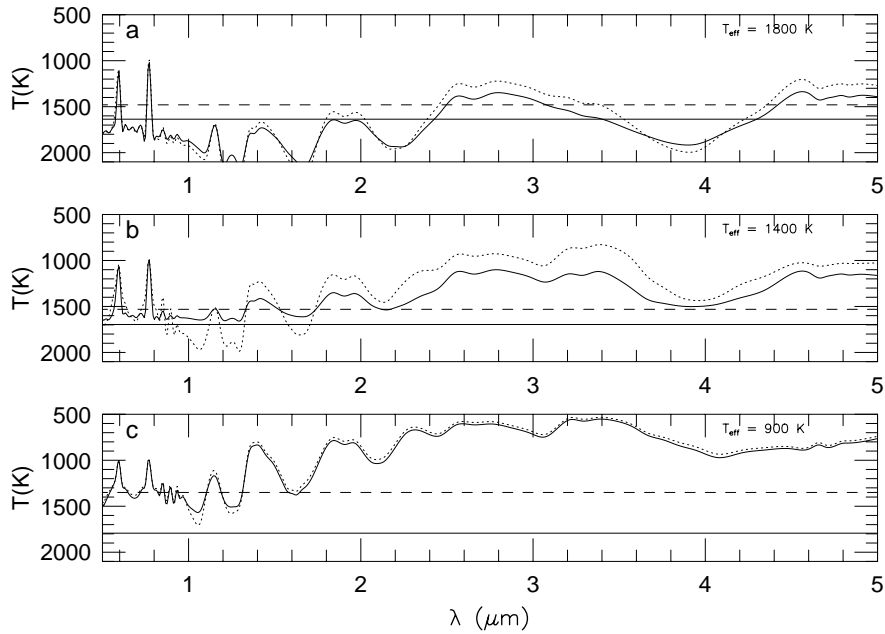}
\caption{Model brightness temperature spectra from Ackerman \&
Marley (2001).  Spectra depict approximate depth in the
atmosphere at which emission arises.  Solid curves depict cloudy 
models and dotted curves
cloud-free models with the same $T_{\rm eff}$ (all for
$g=1000\,\rm m\,sec^{-2}$ \& $f_{\rm rain}=3$).  
Horizontal dashed and solid lines demark the level at which
cloud opacity, integrated from above, 
reaches 0.1 and the base of the silicate cloud, respectively.  In the
early-L like model (a) and the T-dwarf like model (c) clouds play
a relatively small role as they are either optically thin (a) or
form below the level at which most emission arises (c).  Only in the late-L case
(b) do the optically-thick clouds substantially alter the emitted
spectrum and limit the depth from which photons emerge.
Cloud base varies with pressure and cloud thickness varies with
strength of convection, accounting for the varying cloud base temperature
and thickness. }
\end{figure}

\section{Clouds and the L- to T-dwarf transition}
Given the importance of clouds to the L-dwarf spectra and
the absence of significant cloud opacity in the T-dwarfs,
it is clear that the departure of clouds with falling
$T_{\rm eff}$ is an important
milestone in the transition from L- to T-dwarfs.  Marley (2000)
demonstrated that a simple cloud model in which the silicate
cloud was always one scale-height thick could account for
the change in $J-K$ color from the red L-dwarfs to the blue T-dwarfs.
Now using the more physically motivated cloud model of Ackerman
\& Marley we can better test this hypothesis.

Figure 3 illustrates the brightness temperature spectra of
six brown dwarf models with three different $T_{\rm eff}$.  In the
warmest and coolest cases ($T_{\rm eff} = 1800$ and 900 K) models
with and without clouds appear similar.  In the warmer case silicate
and iron clouds are just forming in the atmosphere and are relatively
optically thin, so their influence is slight.  In the cooler case
as in the right-hand image of Uranus in Figure 1, the main cloud
deck forms below the visible atmosphere.  In the intermediate
case ($T_{\rm eff} = 1400\,\rm K$) an optically thick cloud forms in the
visible atmosphere and substantially alters the emitted spectrum.
The atmospheric structure predicted by the Ackerman \& Marley (2001)
model for this case is similar to that inferred by Basri et al. (2000)
from Cs line shapes in L-dwarf atmospheres.
Thus a cooling brown dwarf moves from relatively cloud free conditions
to cloudy to clear.

The solid lines in Figure 2 show how the $J-K$ color evolves
with $T_{\rm eff}$.  Objects first become red as dust begins to
dominate the visible atmosphere, then blue as water and methane
begin to absorb strongly in K band.  Models in which the dust does
not settle (Chabrier et al. 2000) predict $J-K$ colors much redder
than observed.  Instead the colors of
the L-dwarfs are best fit by models which include some precipitation
as parameterized by $f_{\rm rain}= 3$ to 5.  
The data clearly require models for objects cooler than the latest L-dwarfs
to rapidly change from $J-K\sim 2$ to 0 over a relatively small 
$T_{\rm eff}$ range.  While models with $f_{\rm rain} = 3$ to 5 
do turn blue as the clouds sink below the visible
atmosphere (Figure 2), the variation is not rapid enough to satisfy
the observational constraints.
Ackerman \& Marley suggest that holes in the clouds may
begin to dominate the disk-averaged spectra as the clouds are
sinking out of sight.  Jupiter's 5-$\rm \mu m$ spectrum is 
indeed dominated by flux emerging through holes in its clouds.
Bailer-Jones \& Mundt (2000) find variability in L-dwarf atmospheres
that may be related to such horizontal cloud patchiness.

Despite the successes of the Ackerman \& Marley model, clearly much
more work needs to be done to understand clouds in the brown dwarfs.
Perhaps three dimensional models of convection coupled to radiative
transport will be required.

\section{Extrasolar Giant Planets}

The issues of cloud physics considered above of course will also
apply to the extrasolar giant planets (Marley 1998; Marley et al. 1999;
Seager, Whitney, \& Sasselov 2000; Sudarsky, Burrows, \& Pinto 2000).
These papers demonstrate that the reflected spectra of extrasolar
giant planets depends sensitively on the cloud particle size and
vertical distribution.  As already demonstrated by the brown dwarfs
in the foregoing section, the emergent thermal flux is similarly affected.

Indeed Sudarsky et al. suggest that a classification scheme based
on the presence or absence of specific cloud layers be used to
categorize the extrasolar giant planets.  Moderate spectral resolution
transit observations of close-in EGPs, if the bandpasses
are correctly chosen, will certainly provide first-order information
on cloud heights and vertical profiles of these atmospheres (Seager \&
Sasselov 2000; Hubbard et al. 2001).  Coronagraphic multi-wavelength
imaging of extrasolar giant planets will provide similar information
(see Figure 1).

\section{Conclusion}

It is ironic that although the physics governing the vast bulk of the
mass of brown dwarfs and extrasolar planets is very well in hand, 
the old problem of weather prediction governs the radiative
transfer and thus the only remotely sensed quantity.  The good news
is that there will soon be much more weather to talk about, even
if we aren't any farther along in doing anything about it.

\acknowledgments
This work was supported by NASA grant NAG5-8919 and NSF grants AST-9624878 and
AST-0086288.  The authors benefited
from conversations with Dave Stevenson, Sara Seager, 
Adam Burrows, and Bill Hubbard.  Heidi Hammel and Kevin Zahnle offered 
particularly helpful comments on an earlier draft of this contribution.

\end{document}